\documentclass{icrc2009}

\usepackage{graphicx}   
\usepackage{caption}    
\usepackage[font=footnotesize]{subfig} 
\usepackage{fixltx2e}
\usepackage{url}

\newcommand{\shorttitle}[1]%
{\markboth{Proceedings of the 31\MakeLowercase{$^{st}$} ICRC, {\L}\'{o}d\'{z} 2009}{#1} }
\newcommand{\etal}{\MakeLowercase{\textit{et al. }}} 

\begin{document}
\title{Calibrating laser test-beams for cosmic-ray observatories }

\author{\IEEEauthorblockN{Lawrence Wiencke \IEEEauthorrefmark{1},
                          Fernando Arqueros\IEEEauthorrefmark{2},
                           John Compton\IEEEauthorrefmark{1},\\
                           Maria Monasor\IEEEauthorrefmark{2},
                          David Pilger\IEEEauthorrefmark{1} and
                          Jaime Rosado\IEEEauthorrefmark{2}
}
                            \\
\IEEEauthorblockA{\IEEEauthorrefmark{1}Colorado School of Mines, Golden Colorado 80401}
\IEEEauthorblockA{\IEEEauthorrefmark{2}Universidad Complutense de Madrid. Madrid 28040. Spain}

}

\shorttitle{Wiencke \etal Laser Calibration}
\maketitle

\begin{abstract} 
Pulsed UV lasers can provide useful "test-beams" for observatories that use optical detectors, especially fluorescence detectors, to measure high energy cosmic-rays. The light observed by the detector is proportional to the energy of the laser pulse. Since the absolute laser energy can be measured locally, a well-calibrated laser offers a practical way to test the photometric calibration of the cosmic-ray detector including atmospheric corrections. This poster will describe a robotic system for laser polarization and energy calibration. Laboratory measurements of laser energies and polarizations by energy probes from different manufactures will be presented.
  \end{abstract}

 
\section{Introduction}
Modern ultra high energy cosmic-ray experiments base their absolute energy scales on fluorescence detector (FD) measurements. One compelling motivation for minimizing systematics in the absolute energy scale is the strong energy dependence of the propagation distance for cosmic-rays through the cosmic microwave background radiation. For example, a  $25\%$ decrease in energy scale from $8\times10^{19}$ eV to $6\times10^{19}$ eV corresponds to a 10-fold increase of observable volume, or possible source regions.
  
\section{Lasers as Test-Beams}
Pulsed UV lasers fired into the aperture of an FD can generate optical signatures with similarities to a trans-GZK energy (${E>6x10^{19}eV}$) extensive air-shower (EAS) \cite{clfjinst}. Light scattered out of the laser beam by the atmosphere produces a track in the same detectors that also measure tracks generated by cosmic rays.

In optical equivalence, a 5 mJ per pulse laser corresponds roughly to an EAS of ${10^{20}eV}$.  Laser induced tracks tend to be longer than EAS induced tracks. While both depend on the density profile of the atmosphere, different processes are involved. Laser light propagation depends on atmospheric scattering lengths that are typically 10-30 km, depending on optical clarity and atmospheric depth. In contrast, EAS development depends on particle interaction lengths that are much shorter.  Although they travel in the opposite direction of EASs, the tracks from ground-based lasers appear brighter near the ground and dimmer at higher elevations as do the tracks from trans-GZK energy EASs of modest zenith angle.

Past experiments (Fly's Eye and HiRes) relied and present experiments (Pierre Auger (South) and Telescope Array) do rely on atmospheric laser test-beams. In the simplest terms, if the detector in question records and reconstructs distant laser shots well during nightly operations, there is a reasonable expectation that the same detector repeats this feat for the much rarer trans-GZK energy cosmic-rays. This conceptually simple diagnostic will be especially useful for future projects. These include the Pierre Auger Observatory's Northern detector, and the JEM-EUSO and OWL space-based instruments. All plan to monitor much larger volumes of atmosphere from greater distances. 

An especially powerful test is to compare the energy of the laser pulse as measured in the laser enclosure with the energy of the laser as reconstructed from the track it produces in the FD after applying the geometric and most of the same calibration corrections used for cosmic-rays \cite{clfjinst} \cite{Wiencke2009}. Typical uncertainties in FD energy calibration including atmospheric effects currently fall in the 20\% range \cite{Abbasi2008} \cite{Abram2008}.  An apparatus capable of establishing and maintaining the laser energy calibration to a significantly smaller uncertainty (between 5\% and 10\%) over the multi-year duration of the project is therefore desirable.

 \begin{figure}
  \centering
  \includegraphics[width=2.8in]{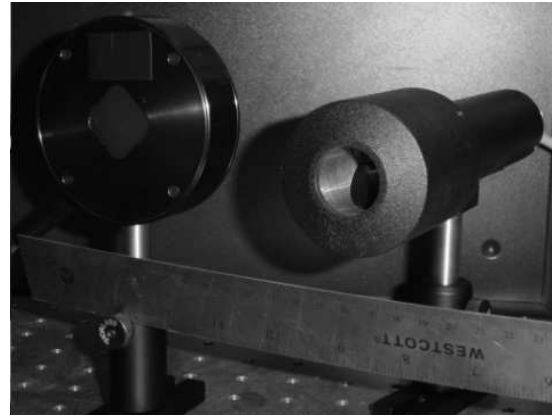}
  \caption{Examples of the two pyroelectric energy probes tested. (Ophir PE-25BB left) (Laserprobe RjP-734 right)}
  \label{probe_fig}
 \end{figure}

 \begin{figure*}[th]
\centering
  \includegraphics[width=4.0in]{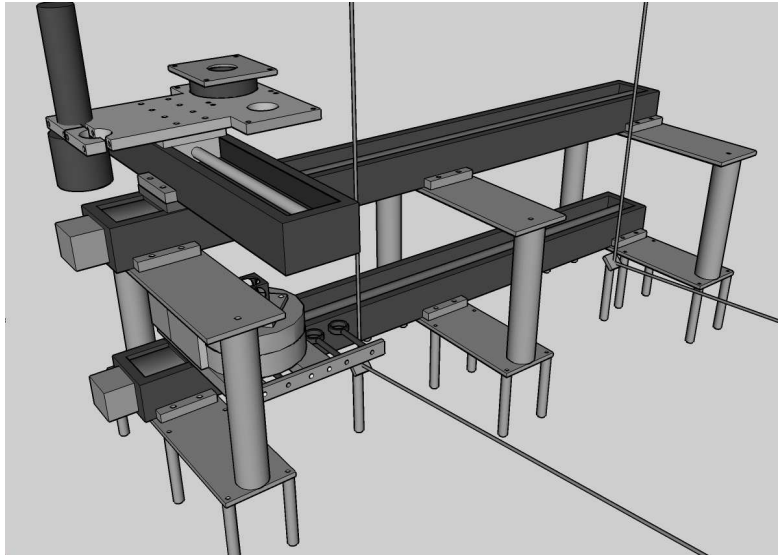}
  \caption{Robotic system for calibrating the energy and 
polarization of two laser beams.}
  \label{xlf-diag}
 \end{figure*}

\begin{figure}
\centering
  \includegraphics[width=2.5in]{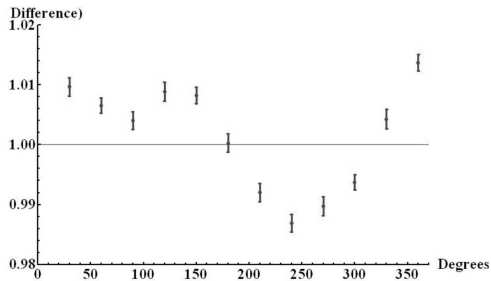}
\caption{Relative response of the Laserprobe RjP-734 probe as a function of its rotation angle about the beam axis.}
\label{LP_rot_fig}
\end{figure}

\begin{figure}
\centering

  \includegraphics[width=2.5in]{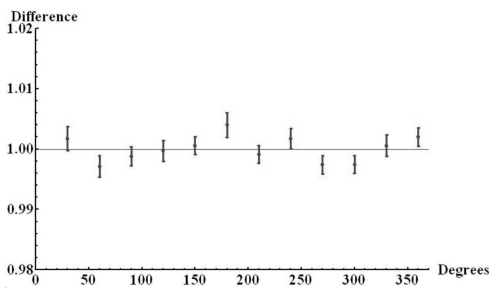}
\caption{Relative response of the Ophir PE-25BB probe as a function of its rotation angle about the beam axis.}
\label{Ophir_rot_fig}
\end{figure}

\section{Lasers and Energy Probes}
Off the shelf frequency tripled 355 nm YAG lasers produce light in the middle of the air fluorescence spectrum. Typical pulse lengths are 7 ns. Diode pumped systems are superior over the traditional liquid cooled flashlamp pumped models because of their longer lifetime and reduced maintenance requirements.

The laser used in these tests was a Quantel Centurion which is the solid state version of the Big Sky CFR Ultra. (Both models are used at Auger.  The latter was used at HiRes and is used at TA.) The divergence of the Centurion's beam was reduced below 1 mR by a downstream beam expander (3X or 5X). This expander reduced the energy density in the beam spot and increased the dynamic range over which the energy probes could be evaluated without exceeding their damage thresholds.

Six pyroelectric energy probes (Figure \ref{probe_fig} and Table \ref{probes_table}) were tested; three from Ophir \cite{ophir-web} model PE-25BB and three from Laserprobe (LP) \cite{laserprobe-web} model RjP-734. The expanded beam-spots of 1.0 to 1.5 cm fit within the probes' 25 mm apertures.

The Ophir probe has a flat absorbing surface while the LP probe has an asymmetric vee shaped cavity designed to reduce its net reflectivity to much less than 1\%.  Initial tests found a non-uniform response at the $\pm1\%$ level (Figure \ref{LP_rot_fig}) depending on the rotation angle of probe along the beam axis.  The effect could be described as a sensitivity to asymmetries the laser's beam profile. In the same test applied to the Ophir probe this behavior was not observed (Figure \ref{Ophir_rot_fig}). 

The quoted damage threshold for the Opher probe is 0.3 ${J/cm^{2}}$. For the LP probe a maximum energy density of 0.4 ${J/cm^{2}}$ and a maximum peak pulse power density of 1.0 ${MW/cm^{2}}$ for a 30 ns pulse are specified. For the 7 ns pulse width of the Centurion laser this corresponds to 0.07 ${J/cm^{2}}$. The damage thresholds were not tested intentionally, however a 5\% decrease in response was observed for two of the LP probes after they were used to measure a few hundred 7 mJ 1.5 cm diameter pulses. (The measurements reported in section V were made prior to this.) The corresponding average energy density of 0.004 ${J/cm^{2}}$ is more than a factor of 10 below the quoted damage threshold. However, beam hot spot effects can not be ruled out as a factor.  

\begin{table}
\centering
\caption{\label{probes_table}Pyroelectric energy probes evaluated (Jan-Aug 2008). Manufacturers' energy calibrations dates are listed in the last column.}
\begin{tabular}{lllll}
\hline\hline

$\textbf{Probe}$&$\textbf{Company}$&$\textbf{Model}$&$\textbf{Serial Number}$&$\textbf{Last Calib.}$\\
\hline
$\textrm{1}$&$\textrm{Laser Probe}$&$\textrm{RjP-734}$&$\textrm{041-074-003}$&$\textrm{02/22/08}$\\
$\textrm{2}$&$\textrm{Laser Probe}$&$\textrm{RjP-734}$&$\textrm{042-074-004}$&$\textrm{02/22/08}$\\
$\textrm{3}$&$\textrm{Laser Probe}$&$\textrm{RjP-734}$&$\textrm{038-074-006}$&$\textrm{04/25/08}$\\
$\textrm{4}$&$\textrm{Ophir}$&$\textrm{PE25BB}$&$\textrm{505291}$&$\textrm{07/14/08}$\\
$\textrm{5}$&$\textrm{Ophir}$&$\textrm{PE25BB}$&$\textrm{221377}$&$\textrm{07/14/08}$\\
$\textrm{6}$&$\textrm{Ophir}$&$\textrm{PE25BB}$&$\textrm{523121}$&$\textrm{03/15/08}$\\
\hline
\end{tabular}
\end{table}

 \begin{figure*}[!t]
 \centerline{\subfloat[PE-25BB]
{\includegraphics[width=2.7in]{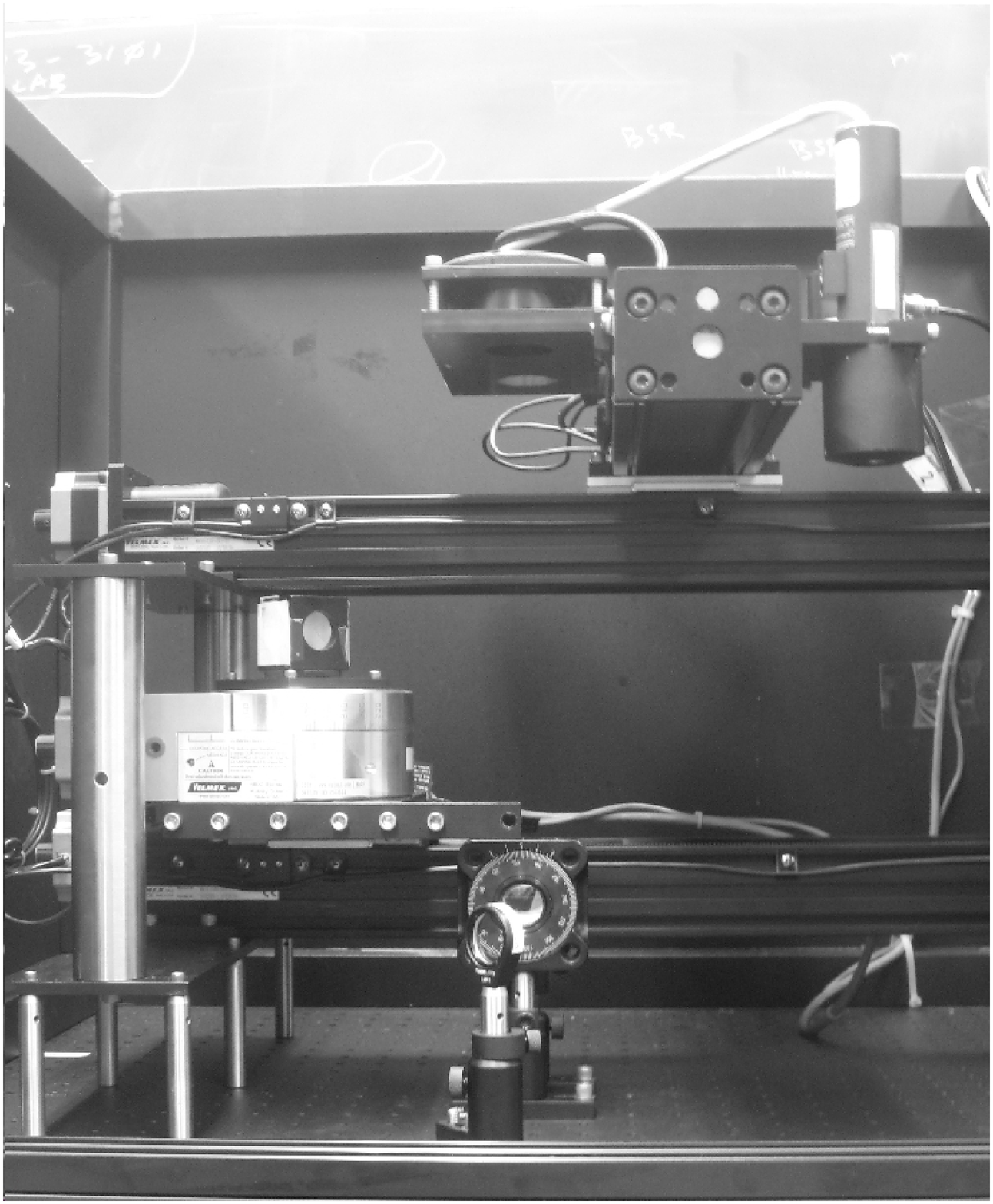} \label{sub_fig1}}
              \hfil
              \subfloat[RjP-734 probe]
{\includegraphics[width=2.7in]{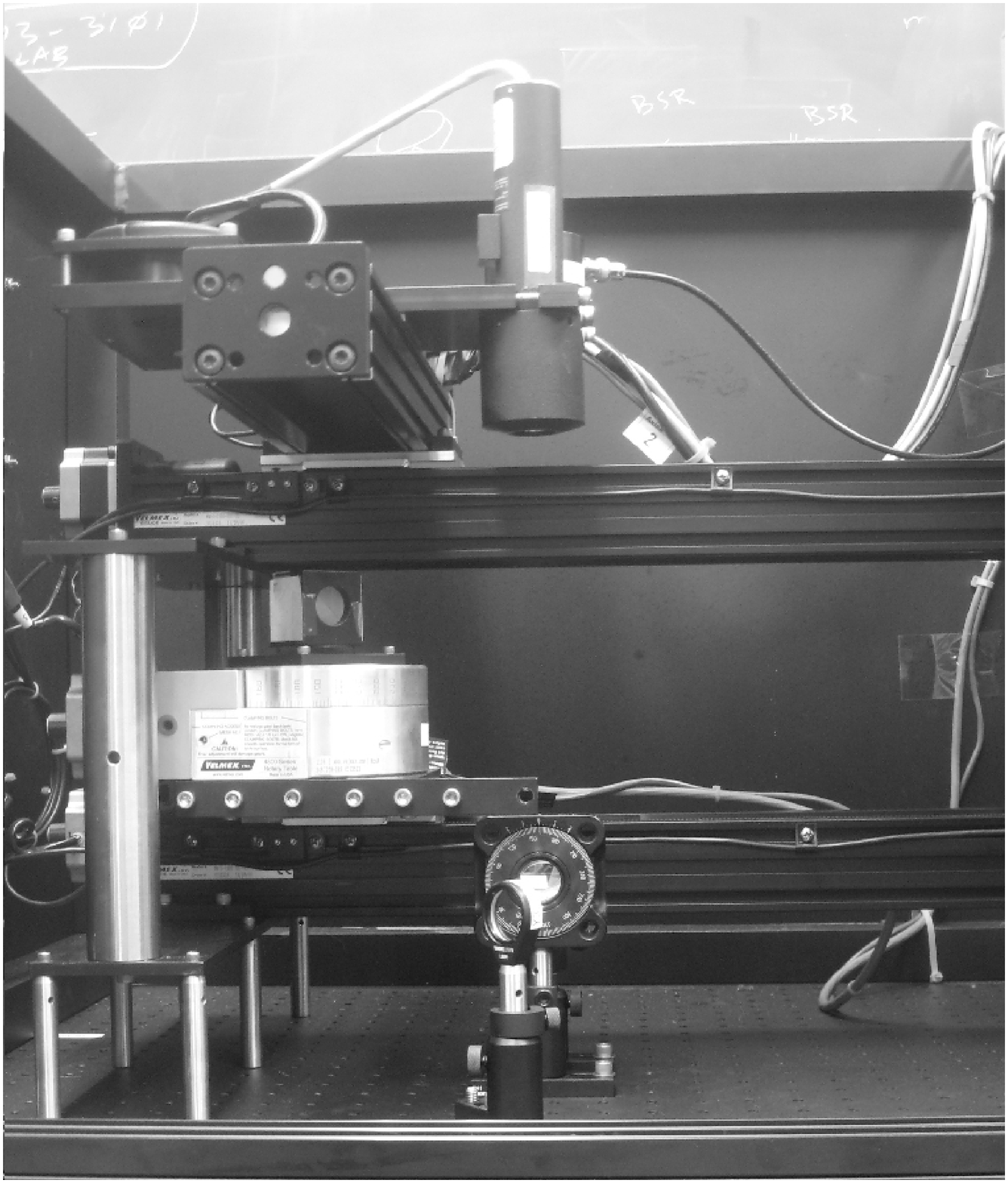} \label{sub_fig2}}
             }
   \caption{Configurations of robotic calibration system used to compare the response of different energy probes to the same vertical laser beam. The rotation stage and polarization analyzer have been moved to the left out of the beam. To produce the polarized beam analyzed in figure \ref{pol_fig} the analyzer is moved into the beam and the energy probes are moved aside}           
   \label{probe_meas}
 \end{figure*}

 \begin{figure*}
 \centerline{\subfloat[1 mJ]
{\includegraphics[width=2.8in]{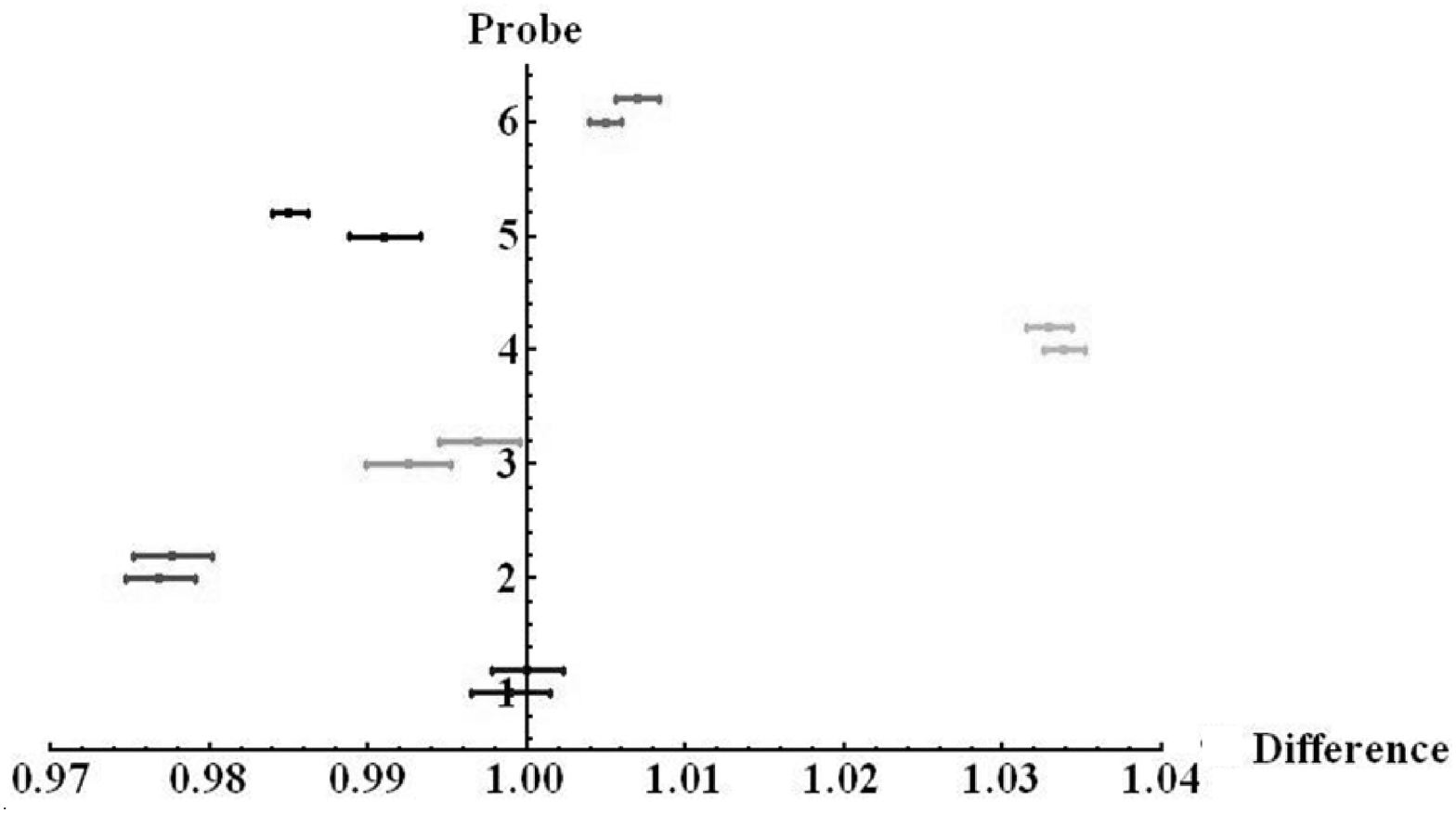}
\label{sub_1mj}} 
\hfil
\subfloat[4.9 mJ] {\includegraphics[width=2.8in]{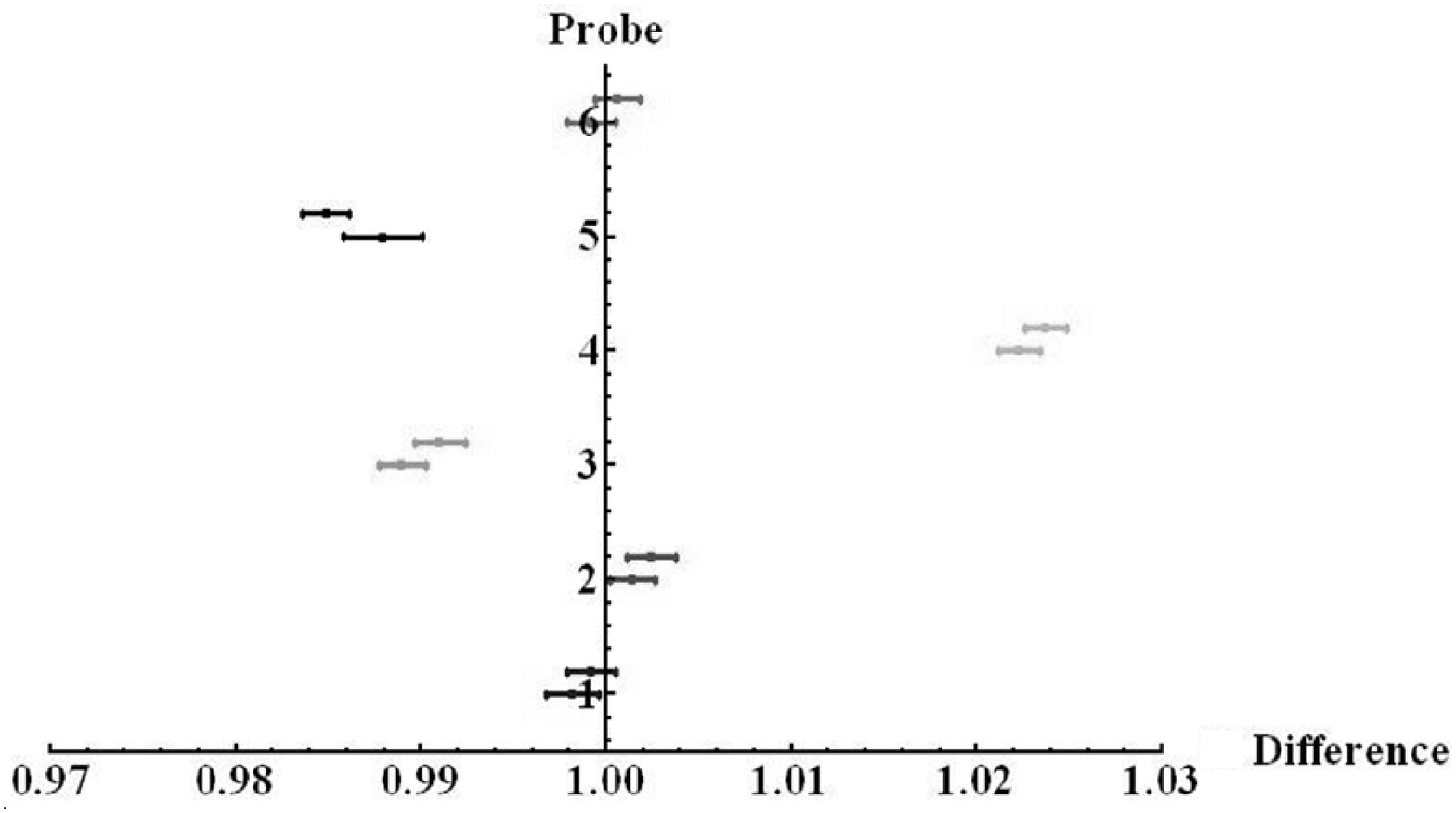}
  \label{sub_4mj}}
}
 \caption{Energy response comparisons between the 6 probes tested for nominal laser energies of 1 mJ (a) and 4.9 mJ (b). The vertical axis corresponds to the probe number (table \ref{probes_table}). 1.00 in on the horizontal is the average of the six measurements for the same nominal energy. Error bars are statistical.  Data pairs correspond two sets of measurements under the same configuration.}
\label{Energy_Meas}
 \end{figure*}
 \noindent

 \begin{figure*}
   \centerline{\subfloat[Polarization Angle]
{\includegraphics[width=2.8in]{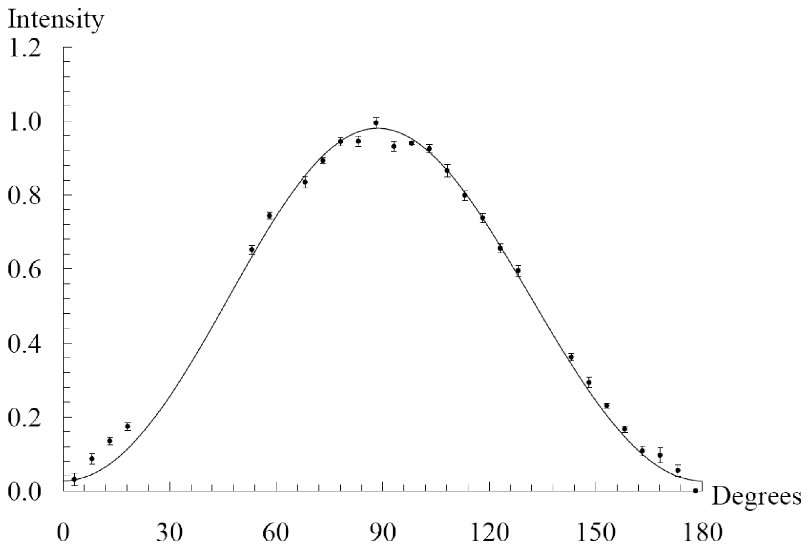} \label{pol_subfig1}}
              \hfil
              \subfloat[Scattering Angle]
{\includegraphics[width=2.8in]{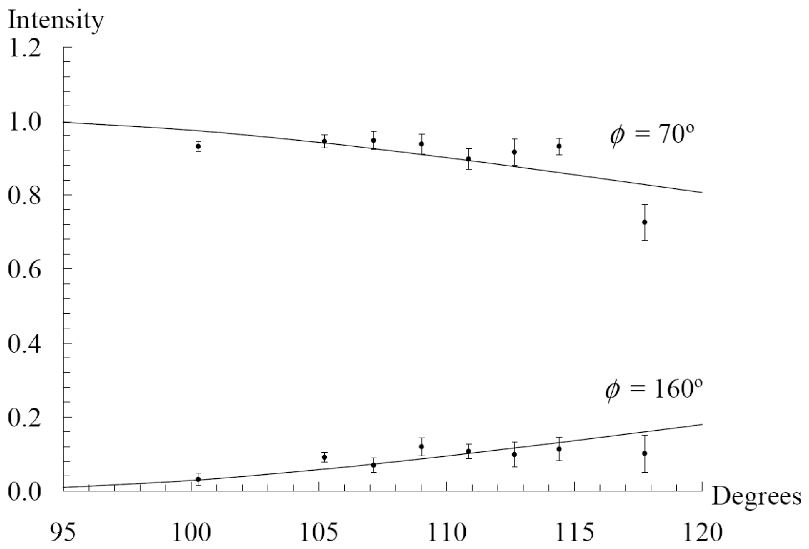} \label{pol_subfig2}}
             }
   \caption{ Intensity of scattered light from vertical linearly polarized beams recorded by a fluorescence telescope normalized to measurements of unpolarized beam under the same atmospheric conditions. Data is shown as a function of the polarization rotation angle of the beam (a) and of the scattering angle (b). The curves show the Rayleigh scattering prediction for polarized light. Error bars correspond only to statistical uncertainties.}           
   \label{pol_fig}
 \end{figure*}
\section{Robotic Laser Beam Calibration}
The robotic calibration system used in these tests was designed for the Pierre Auger (South) eXtreme Laser Facility (XLF). The robotic system can remotely calibrate the energy and polarization of vertical laser beams by moving probes and other components directly in to the beam at a point downstream of the final optical component the beam encounters before heading into the sky. (The XLF has a pick-off energy probe to monitor relative variations in laser energy) Four motorized stages can center equipment on the  vertical laser beams. One linear stage carries a rotary table and two optical filters.  A cube polarizing analyzer is mounted on the rotary table.  Two other linear stages can position a 4-probe bracket in X and Y. (Figure \ref{xlf-diag}).  With this system it is possible to arrange any combination of four different energy probes, and a polarizer or a filter over each of the two vertical beams.  Two combinations are shown in figure \ref{probe_meas}.

\section{Characterization of Energy Probes}
To evaluate manufacturers' specifications of energy calibration, six probes from two companies were tested (Table \ref{probes_table}). All probes were calibrated by the manufactures a few months prior to the laboratory measurements described here. Energy calibration uncertainty specified by Ophir for the probes tested is $\pm3\%$ with additional errors of $\pm2\%$ for wavelength and $\pm1\%$ for frequency. The quoted energy uncertainty for the LP probe is $\pm5\%$. 

The probes were compared against each other to measure their consistency in energy response.  Results for two nominal laser energies of 1.0 and 4.9 mJ are shown in figure \ref{Energy_Meas}. Differences are expressed as a percent variation from the 6 probe average for a given laser energy. For both energies, all measurements fall within the possible range of uncertainties of the manufacturers' calibrations. At 4.9 mJ, the average of the 3 Ophir probes was 1\% higher than that of the 3 LP probe. This number increased to 2\% at 1.0 mJ.  The largest difference between two probes was 5.5\% at 1 mJ and 3.5\% at 4.9 mJ. We note these measurements were made in laboratory conditions over a relatively short period.

\section{Beam Polarization}
A beam of zero net polarization (randomized by a depolarizer) is especially useful for reasons of symmetry. The same amount of light is scattered axially about the beam. Following the techniques described in \cite{clfjinst} de-polarization below 3\% can be achieved. However, since the robotic system can be used to measure polarization it can also be used to produce and rotate a linearly polarized beam from the incident de-polarized beam. To do this the polarization analyzer is left in the beam and all energy probes are moved aside.

Linearly polarized beams can provide an additional test for an FD, since the intensity of scattered light reaching a fluorescence telescope depends on the beam polarization relative to the observation plane (i.e., that defined by the laser line and the position of the actual fluorescence telescope). For instance, Rayleigh scattering in the observation plane vanishes at a scattering angle $\theta=90^\circ$ for light polarized perpendicular to the same plane, and thus, only the parallel polarization component contributes to the scattered intensity at that angle. Mie scattering also depends on the state of polarization as well as other parameters, but it is strongly peaked at small angles. Therefore, under reasonably clear conditions, Mie scattering is negligible as compared with Rayleigh scattering angles around or above $90^\circ$, which typically range the aperture of a fluorescence telescope for vertical beams.

Figure 7a shows the intensity of scattered light from linearly polarized vertical test-beams relative to that from non-polarized ones recorded by a fluorescence telescope as a function of the beam's polarization angle $\phi$. As expected for a scattering angle around $100^\circ$, intensity has an almost pure $\sin^2 \phi$ modulation as corresponds to the parallel polarization component of the beam. For larger scattering angles, perpendicular polarization also contributes to the total observed intensity, and thus, that $\phi$ dependence is smoothed. Accordingly, intensities for two reference polarization angles -around the maximum and minimum, respectively- approach each other as the scattering angle increases (figure 7b).  For a larger range of scattering angles, this offers a way to vary the longitudinal profile in a test-beam track.

\end{document}